\documentclass[12pt,a4paper]{article}
\usepackage{graphicx,epstopdf,setspace}

\usepackage{hyperref}

\usepackage{color}
\usepackage{amsthm,amsfonts,amsmath,amssymb,bbm,color}

\usepackage{cancel}
\usepackage[normalem]{ulem}

\usepackage{lineno}

\makeatletter
\renewcommand{\@listI}{\setlength{\topsep}{4pt}
\setlength{\itemsep}{-1pt} \setlength{\parskip}{0pt}}
\renewcommand{\@listii}{\setlength{\topsep}{-1pt}\setlength{\itemsep}{-3pt}
}

\def\@biblabel#1{\@ifnotempty{#1}{#1}}

\makeatother
\newtheorem{assumption}{Assumption}
\theoremstyle{remark}

\newcommand{\GPD}{\mathrm{GPD}}
\newcommand{\GEV}{\mathrm{GEV}}
\newcommand{\logit}{\mathop{\mathrm{logit}}\nolimits}
\newcommand{\PP}{\mathsf{P}}

\newcommand{\myp}{\mbox{$\:\!$}}
\newcommand{\mypp}{\mbox{$\;\!$}}
\newcommand{\myn}{\mbox{$\:\!\!$}}

\begin{document}


\title{Nonstationary POT
modelling of air pollution concentrations: Statistical analysis of
the traffic and meteorological impact }


\author{J\'anos Gyarmati-Szab\'o\myp$^{\text{1}}$, Leonid V.~Bogachev\myp$^{\text{2}}$, and Haibo Chen\myp$^{\text{3}}$}


\date{\small $^{\,\text{1}}$\myp Institute for Transport Studies and Department of Statistics,
University of Leeds,\\ Leeds LS2 9JT, UK. E-mail: {\tt j.gyarmati.szabo@gmail.com}\\[.2pc]
$^{\,\text{2}}$\myp Department of Statistics, School of Mathematics,
University of Leeds,\\ Leeds LS2 9JT, UK. E-mail: {\tt L.V.Bogachev@leeds.ac.uk}\\[.2pc]
$^{\,\text{3}}$\myp Institute for Transport Studies, University of
Leeds, Leeds LS2 9JT, UK.\\
E-mail: {\tt H.Chen@its.leeds.ac.uk}}

\maketitle

\begin{abstract}
Predicting the occurrence, level and duration of high air pollution
concentrations exceeding a given critical level enables researchers
to study the health impact of road traffic on local air quality and
to inform public policy action. Precise estimates of the
probabilities of occurrence and level of extreme concentrations are
formidable due to the combination of complex physical and chemical
processes involved. This underpins the need for developing
sophisticated extreme value models, in particular allowing for
non-stationarity of environmental time series. In this paper,
extremes of nitrogen oxide ($\text{NO}$), nitrogen dioxide
($\text{NO}_{2}$) and ozone ($\text{O}_{3}$) concentrations are
investigated using two models. Model~I is based on an extended
peaks-over-threshold (POT) approach developed by A.\,C.\,Davison and
R.\,L.\,Smith, whereby the parameters of the underlying generalized
Pareto distribution (GPD) are treated as functions of covariates
(i.e., traffic and meteorological factors). The new Model~II
resolves the lack of threshold stability in the Davison--Smith model
by constructing a special functional form for the GPD parameters.
For each of the models, the effects of traffic and meteorological
factors on the frequency and size of extreme values are estimated
using Markov chain Monte Carlo methods. Finally, appropriate
goodness-of-fit tests and model selection criteria confirm that
Model~II significantly outperforms Model~I in estimation and
forecasting of extremes.

\vspace{.4pc}\noindent \emph{Keywords}\/: Roadside air pollution;
Extreme values; Peaks-over-threshold (POT); Generalized Pareto
distribution (GPD); Non-stationarity; Threshold stability.

\vspace{.4pc}\noindent\emph{2010 MSC}:
Primary: 62G32; Secondary: 60G70, 62F15, 62G05. \hfill{}


\end{abstract}

\section{Introduction}
\label{intro}

The issue of high episodic concentrations of air pollutants (e.g.,
nitrogen oxides, particulates, and carbon monoxide) is a matter of
growing worldwide concern due to their harmful effects on human
health and the environment. Among many contributing factors, road
traffic emission is accounted for high proportions of harmful
pollutants. Accurate prediction of pollution episodes, including
their magnitude and duration, is a formidable problem due to the
combination of many complex physical and chemical processes involved
in their formation. Hence, there is a need for the development of
sophisticated extreme value models in order to facilitate prediction
of high pollution concentrations and to better understand their
cause.

Statistical analysis of air pollution data has been extensively
advanced in the recent decade (see, e.g., Carslaw et al., 2007; Zito
et al., 2008; and references therein). In general, research in this
area aims to pursue the following long-term goals (Thompson et al.,
2001; Eastoe \& Tawn, 2009):
\begin{itemize}
\item
predicting critical levels of pollutants to give out health warnings
to public;

\item
identifying and predicting trends in high concentration levels;

\item
assessing changes in air pollution levels due to the impact of human
activities on the environment, either direct (e.g., via changing
emission patterns) or indirect (through the climate change).
\end{itemize}

As argued by Eastoe and Tawn (2009), statistical methodology based
on extreme value theory is particularly suited to address these
problems. An extensive discussion of extreme value methods useful in
the air quality modelling can also be found in Horowitz and Barakat
(1979), Smith (1989), and K\"uchenhoff and Thamerus (1996).

In the present paper, a novel approach to extreme value modelling is
proposed, tailored to the problems indicated above on both long
(e.g., daily or yearly) and short (e.g., 15 min) time scales. The
latter is particularly important in environmental and health
applications due to the fact that even a short exposure to high
pollution concentrations may have harmful effects on human health,
for example, asthmatic patients exposed to a high sulphur dioxide
($\text{SO}_{2}$) concentration may develop adverse symptoms within
minutes (WHO, 2000).

\begin{figure}
\centering
\includegraphics[width=1.00\textwidth]
{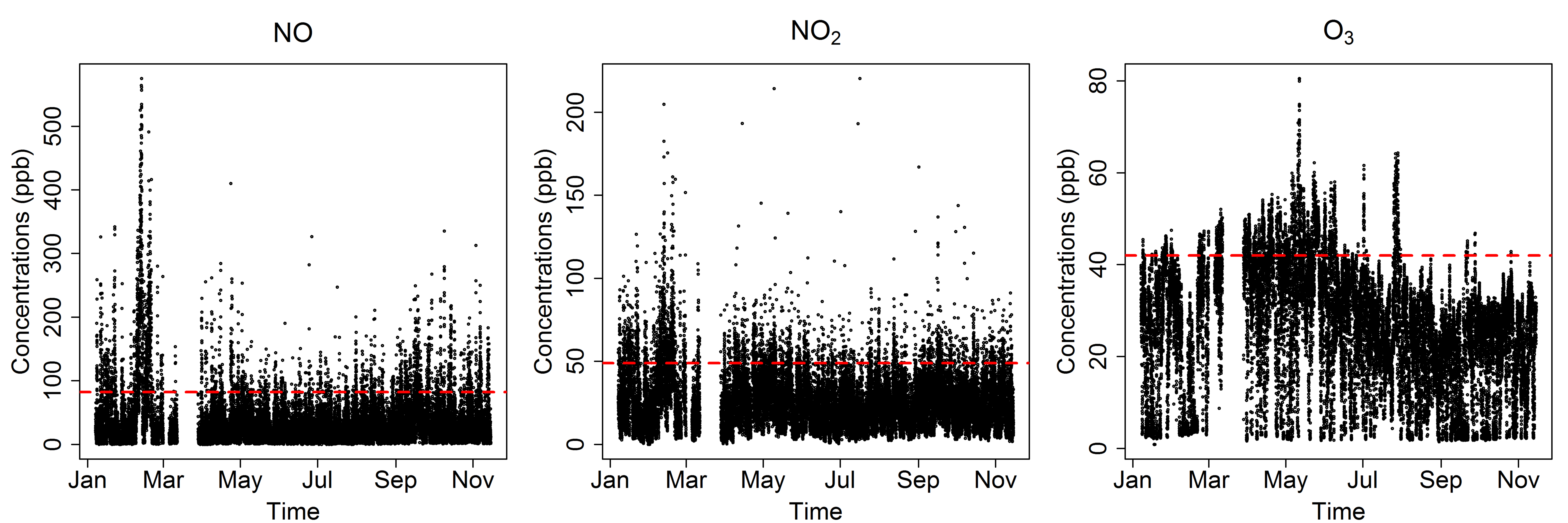} \caption{Time-series plots of 15-min maximum
concentrations of $\text{NO}_{x}$ and $\text{O}_{3}$ (expressed in
ppb = parts-per-billion), collected at the Kirkstall roadside
laboratory from 1 January 2008 till 1 January 2009. Missing data are
left blank. The empirical $90\%$-quantiles of the concentration
values are indicated by the dashed lines (red in the online
version).} \label{fig:Time_Series_concentration}
\end{figure}

The models presented in this paper are used to analyse the nitrogen
oxide ($\text{NO}$), nitrogen dioxide ($\text{NO}_{2}$),
collectively referred to as $\text{NO}_{x}$, and ozone
($\text{O}_{3}$) concentration data (see
Figure~\ref{fig:Time_Series_concentration}), which consist of
$15$-min maxima of $1$-min concentration values observed within one
calendar year, from January 1, 2008 to January 1, 2009. The data
were collected at a fixed roadside laboratory on Kirkstall Road in
the city of Leeds (West Yorkshire, UK). The laboratory houses a
traffic monitoring system and an air quality monitoring station, in
which local meteorological conditions are recorded (see more details
in Section~\ref{subsec:Covariates}).

As demonstrated in Figure~\ref{fig:Time_Series_concentration}, all
three chemicals have a highly variable dynamic behaviour, with
noticeable evidence of nonstationarity. Temporal variation in the
mean and variance, as well as in the patterns of extreme values, can
be explained by the underlying physical and chemical processes. An
additional remarkable feature of these plots is that the three
chemicals are apparently strongly correlated, which can be
attributed to the photochemical interconversion reactions between
the pollutants (Clapp \& Jenkin, 2001). Despite a large number of
chemical processes potentially oxidizing $\text{NO}$ and
$\text{NO}_{2}$, the most important source of secondary
$\text{NO}_{2}$ is the following reaction:
\begin{equation}\label{Eq:1}
\text{NO} + \text{O}_3 \to \text{NO}_2 + \text{O}_2,\qquad
\text{NO}_2 + h\nu \,(+\text{O}_2) \to \text{NO} + \text{O}_3.
\end{equation}
Due to the asymmetry of the reactions \eqref{Eq:1}, the equilibrium
concentration levels (photostationary state) strongly depend on the
presence (or otherwise) of sunlight ($h\nu$), in particular leading
to decreased levels of $\text{NO}_{x}$ (and, correspondingly,
increased levels of ozone $\text{O}_{3}$) during the summer due to a
higher solar radiation. Hence, there are yearly cycles in the time
series of concentrations. There is also a difference between
daylight and night-time measurements that can be explained by the
lack of sunshine at night and dry deposition in the early morning.
In addition, the emission of $\text{NO}_{x}$ also follows some
daily, weekly and seasonal trends, which are due to the traffic
patterns (including peak hours and weekday--weekend oscillations)
and to the seasonal variability of human activities, such as
emission from plants and motor vehicles, combustion of fossil fuels
in habitations, etc. Specifically, the data under study correspond
to traffic-dominated emission sources.

In the present work, the extreme value models are developed
furnishing an efficient tool to estimate the probabilities of future
extreme events, conditional on the traffic and meteorological
covariates. In particular, these models account for nonstationarity
of pollution concentrations; thus the effects of possible future
scenarios relating to changes in emission patterns and climate can
also be quantified. Despite the apparently high correlation between
the three pollutants under study, in this work we confine ourselves
to univariate models for each chemical, which however prove to be
rather efficient (see further comments in
Section~\ref{sec:TSummary}).

\section{Peaks-over-threshold models}

Suppose that observations $X_1,X_2\dots$ are independent and
identically distributed (i.i.d.)\ random variables with a common
continuous distribution function $F(x)=\PP(X_t\le x)$
($-\infty<x<\infty$). There are two main approaches in the
literature to characterization of extreme values in the sequence
$(X_t)$ (see Smith, 1989, for a review and further bibliography
therein). The classic \emph{block-maxima} method based on the
Fisher--Tippett--Gnedenko theorem (see Embrechts et al., 1997)
proceeds by picking the maximum values over certain blocks of data
of large enough size $n$ (e.g., one year worth of data each) and
approximating them via a \emph{generalized extreme value (GEV)}
distribution function,
\begin{equation}\label{Eq:GEV}
\PP\bigl(\max\{X_1,\dots,X_n\} \le x\bigr)\approx
F_\GEV(x;\mu,\sigma,\xi)=\exp\left\{-
\left(1+\frac{\xi(x-\mu)}{\sigma}\right)_{+}^{-1/\xi}\right\},
\end{equation}
where $a_+=\max\{a,0\}$ and $\mu$, $\sigma$ and $\xi$ are the
\emph{location}, \emph{scale} and \emph{shape} parameters,
respectively. Three distinct subclasses of the GEV family
\eqref{Eq:GEV} are known as the \emph{Fr\'echet} ($\xi>0)$,
\emph{Weibull} ($\xi<0$), and \emph{Gumbel} ($\xi=0$) distributions
(the latter is interpreted as the limit of \eqref{Eq:GEV} as $\xi
\rightarrow 0$).

A more recent \emph{peaks-over-threshold (POT)} method (see
Embrechts et al., 1997) focuses on exceedances within the original
sample $(X_t)$ over a high threshold $u$, that is, $X_t^u=X_t-u$
conditioned on the event $X_t>u$. The phrase ``high threshold''
means that $u$ is close to the upper boundary of the support of the
GEV distribution \eqref{Eq:GEV}, that is, $u\to\mu+\sigma/(-\xi)$
($\xi<0$) or $u\to\infty$ ($\xi\ge0$). In particular, the
\emph{threshold exceedance rate}
\begin{equation}\label{eq:rho}
\rho_u= \PP(X_t>u)
\end{equation}
is small. The corresponding approximation, conveniently written for
the tail distribution function, is known as the \emph{generalized
Pareto distribution (GPD)},
\begin{equation}\label{Eq:GPD -Di.F.}
\PP(X_t^u>x)=\PP(X_t> u+x\, |\, X_t>u) \approx
\bar{F}_{\GPD}(x;\sigma_u,\xi)=\left(
1+\frac{x\,\xi}{\sigma_u}\right)_{+}^{-1/\xi},\qquad x\ge 0,
\end{equation}
with the same shape parameter $\xi$ as in \eqref{Eq:GEV} (hence, not
depending on the threshold value~$u$) and a new scale parameter
$\sigma_u=\sigma+\xi(u-\mu)$. The GPD approximation \eqref{Eq:GPD
-Di.F.} holds whenever the GEV approximation \eqref{Eq:GEV} does
(Pickands, 1975). Note that for $\xi\ge0$ the support of the GPD
\eqref{Eq:GPD -Di.F.} is the half-line $[0,\infty)$, whereas for
$\xi<0$ it is a bounded interval, $0\le x\le \sigma_u/(-\xi)$.

Because the air quality standards and objectives are normally
expressed in terms of certain critical thresholds, the POT approach
is apparently more suitable and better interpretable in the air
pollution context. Moreover, there is a growing recognition among
statisticians and environmental modellers that the POT is superior
to the block-maxima approach in that the latter ignores any
secondary extreme values in each block (i.e., next to the largest
one) and thus leads to loss of information; consequently, the POT
has a strong potential to provide an improved accuracy of estimation
and inference (Coles, 2001).

In the influential paper of Davison and Smith (1990), an extension
of the POT method to the \emph{nonstationary case} was proposed,
based on treating the parameters of the GPD as \emph{functions of
covariates}. This method was applied in the air pollution context,
for example, by Eastoe and Tawn (2009), to model the extremes of
surface level ozone ($\text{O}_{3}$) concentrations.

In the present paper, our first goal is to build a dynamic version
of the Davison--Smith model (referred to as \emph{Model~I}) to
estimate the distribution of extreme concentration values (threshold
exceedances) in terms of the traffic and meteorological conditions
and also to assess the impact of these covariates on the extremes.
Furthermore, we propose a special functional form for the parameters
of the GPD (leading to the new \emph{Model~II}) to ensure
consistency over different threshold choices. This resolves the
intrinsic problem of \emph{threshold stability} of the GPD in the
nonstationary case (cf.\ Eastoe \& Tawn, 2009; Northrop et al.,
2016); to the best of our knowledge, such a solution is new. Due to
analytical and computational complexities, estimation of the model
parameters and variable selection in both Models I and~II were
carried out within the Bayesian framework by using a suitable Markov
Chain Monte Carlo (MCMC) procedure (Gilks et al., 1996).

The rest of the paper is organized as follows. The methodology based
on an extension of the Davison--Smith model is set out in
Section~\ref{sec:Methodology}. The results of the model fitting are
presented in Section~\ref{sec:Result} along with the estimation of
extreme events. This is followed in Section~\ref{sec:TSummary} by a
discussion of the inference obtained.

\section{Methodology}
\label{sec:Methodology}

\subsection{GPD model with covariates}
Let us introduce the necessary notation and describe our models in
more detail. Let $(X_t)$ be a (discrete-time) random process
representing the temporal dynamics of the concentration of a
specific pollutant, and let $u$ be a selected (high) threshold
level. If the observed values $(X_t)$ are i.i.d.\ and a GEV
approximation \eqref{Eq:GEV} holds, then the (conditional)
distribution function of the threshold exceedances $X_t^u=X_t - u$
can be approximated by the GPD (see \eqref{Eq:GPD -Di.F.}).

Following Davison and Smith (1990), nonstationarity of observations
$(X_t)$ (and hence of the threshold exceedances) is incorporated in
Model~I through the conditional distribution
\begin{equation}\label{level}
\PP(X^u_t>x\,|\,\boldsymbol{C}_t =\boldsymbol{c})=\PP(X_t>
u+x\,|\,X_t>u, \,\boldsymbol{C}_t =\boldsymbol{c}),\qquad x\ge 0,
\end{equation}
where $\boldsymbol{C}_t$ denotes the random vector of relevant
covariates at time $t$ (such as traffic and meteorological
conditions), with values $\boldsymbol{c}$ in a suitable covariate
space $\mathcal{C}$. Full details of the covariates used in the
model building are described in Section~\ref{subsec:Covariates}. The
marginal distribution of the exceedances $X_t^u$ is obtained by
integrating over the covariates,
\begin{equation}\label{Eq:X_ut-unconditional}
\PP(X^u_{t}>x) = \int_{\mathcal{C}} \PP\left(X^u_{t}> x\,|\,
\boldsymbol{C}_t=\boldsymbol{c}
\right)f_t(\boldsymbol{c})\,\mathrm{d}\boldsymbol{c},\qquad x\ge0,
\end{equation}
where $f_t(\boldsymbol{c})$ is the density of $\boldsymbol{C}_t$. In
practice, the function $f_t(\boldsymbol{c})$ is not known but can be
estimated by its empirical version using a subset of observation
times where the covariate vector takes values close to a given
$\boldsymbol{c}\in\mathcal{C}$.

The statistical analysis is greatly simplified under the standard
assumption that the \emph{conditional distribution \eqref{level}
does not depend on time~$t$}. That is to say, the covariate vector
$\boldsymbol{C}_t$ plays a role of a \emph{sufficient statistic}
incorporating all information contained in the data about
nonstationarity of the threshold exceedances. This assumption is in
fact the backbone of the Davison--Smith modelling approach.

As a result, the GPD approximation \eqref{Eq:GPD -Di.F.} takes the
form
\begin{equation}\label{Eq:GPD - Con.Di.F.}
\PP(X^u_t>x\,|\,\boldsymbol{C}_t=\boldsymbol{c})\approx
\bar{F}_{\GPD} \left(x;\sigma_u(\boldsymbol{c}),\xi(\boldsymbol{c})
\right)=\left( 1+\frac{x\, \xi
(\boldsymbol{c})}{\sigma_u(\boldsymbol{c})}\right)_{+}^{-1/
 \xi(\boldsymbol{c})},\qquad x\ge0,
\end{equation}
where the GPD parameters $\sigma_u=\sigma_u(\boldsymbol{c})$ and
$\xi=\xi(\boldsymbol{c})$ are functions of the current value of the
covariate vector $\boldsymbol{C}_t=\boldsymbol{c}$.

Likewise, the rate of threshold exceedance at time $t$
(see~\eqref{eq:rho}) should also be interpreted as a
covariate-dependent quantity,
\begin{equation}\label{eq:rho_c}
\rho_u(\boldsymbol{c})
=\PP(X_t>u\,|\,\boldsymbol{C}_t=\boldsymbol{c}).
\end{equation}
Note that the probability of the event $X_t>u+x$ conditioned on the
covariates (i.e., the exceedance rate for a  higher threshold $u+x$)
is given by
\begin{align}
\notag
\rho_{u+x}(\boldsymbol{c})&=\PP(X_t>u+x\,|\,\boldsymbol{C}_t=\boldsymbol{c})\\
\notag &= \PP(X_t>u\,|\,\boldsymbol{C}_t=\boldsymbol{c})\cdot
\PP(X_t>u+x\,|\,X_t>u,\,\boldsymbol{C}_t=\boldsymbol{c})\\
\notag &= \rho_u(\boldsymbol{c})\cdot
\PP(X^u_t>x\,|\,\boldsymbol{C}_t=\boldsymbol{c})\\
\label{eq:rhO_u+x}
&\approx\rho_u(\boldsymbol{c}) \left( 1+\frac{x\,
\xi (\boldsymbol{c})}{\sigma_u(\boldsymbol{c})}\right)_{+}^{-1/
\xi(\boldsymbol{c})},
\end{align}
according to GPD \eqref{Eq:GPD - Con.Di.F.}.

Let us now introduce the usual regression-type parametric assumption
(cf.\ Davison \& Smith, 1990; Eastoe, 2009; Eastoe \& Tawn, 2009)
for the functional link between the model parameters and the
covariates. To be specific, suppose that the covariate vector
$\boldsymbol{c}$ is $m$-dimensional (including any dummy variables
if required) and, without loss of generality,
$\boldsymbol{c}\in\mathbb{R}^m$. In order to accommodate baseline
effects (intercepts) in our regression models, it is also convenient
to introduce the \emph{extended covariate vector}
$$
\tilde{\boldsymbol{c}}^\top=(1,c_1,\dots,c_m)\in\mathbb{R}^{m+1},
$$
where $\boldsymbol{x}^\top$ denotes the transpose of a (column)
vector $\boldsymbol{x}$. In what follows, we use the standard
notation
$$
\logit \rho=\ln\left(\frac{\rho}{1-\rho}\right),\qquad 0<\rho<1.
$$

\begin{assumption}\label{as1}
The quantities $\ln \sigma_u (\boldsymbol{c})$,
\,$\xi(\boldsymbol{c})$ and $\logit \rho_{u}(\boldsymbol{c})$ are
linear functions of the covariates,
\begin{align}\label{Eq:param_covariat_functional_form}
\ln \sigma_u(\boldsymbol{c}) =
\boldsymbol{s}_u^{\top}\myp\tilde{\boldsymbol{c}},\ \ \quad
\xi(\boldsymbol{c}) =
\boldsymbol{\kappa}^{\top}\tilde{\boldsymbol{c}},\ \ \quad \logit
\rho_u(\boldsymbol{c}) =
\boldsymbol{r}_u^{\top}\myp\tilde{\boldsymbol{c}},
\end{align}
where $\boldsymbol{s}_u$, $\boldsymbol{\kappa}$ and
$\boldsymbol{r}_u$ are the corresponding $(m+1)$-dimensional vector
coefficients (effects).
\end{assumption}
The regression setting \eqref{Eq:param_covariat_functional_form} is
the basis of our \emph{Model~I}.

\subsection{Threshold stability}
\label{subsec:Threshold Stability}

The desirable property of the GPD is its consistency with regard to
the variable choice of the threshold, referred to as the
\emph{threshold stability}: if the conditional distribution of
exceedances over $u$ is a GPD with parameters $\sigma_u$ and $\xi$,
then the conditional distribution of exceedances over a higher level
$u+x$ ($x\ge0$) should also be given by a GPD with the same shape
parameter $\xi$ and the new scale parameter $\sigma_{u+x}=\sigma_u
+x\,\xi$ (Embrechts et al., 1997). In the nonstationary case, this
conditions transcribes as
\begin{equation}\label{Eq:Thres_stab_nonstat_form}
\sigma_{u+x}(\boldsymbol{c})=\sigma_u(\boldsymbol{c})
+x\,\xi(\boldsymbol{c}).
\end{equation}
As was pointed out by Eastoe and Tawn (2009), the Davison--Smith
model in general does not guarantee the threshold stability: even if
$\xi(\boldsymbol{c})$ is constant, $\sigma_{u+x}(\boldsymbol{c})$
must be either constant (leading back to the stationary case) or a
linear function, which contradicts the log-linear formulation of the
model~\eqref{Eq:param_covariat_functional_form}.

We suggest a solution to this problem by using a special functional
form for the scale and shape parameters, which then leads to what we
call \emph{Model~II}. Observe that, according to
\eqref{Eq:Thres_stab_nonstat_form}, $\sigma_u(\boldsymbol{c})$ is
linear in $u$. Thus, the proposed functional parameterization is as
follows:
\begin{equation}
\label{Eq:Thres_stab_nonstat_form_sigm_PROPOSED}
\sigma_u(\boldsymbol{c})= \bigl(\alpha(\boldsymbol{c}) + u\myp
\beta(\boldsymbol{c})\bigr)\,\mathrm{e}^{\gamma(\boldsymbol{c})},\qquad
\xi(\boldsymbol{c})=\beta(\boldsymbol{c})\,\mathrm{e}^{\gamma(\boldsymbol{c})},
\end{equation}
where $\alpha(\boldsymbol{c})$, $\beta(\boldsymbol{c})$ and
$\gamma(\boldsymbol{c})$ are functions of the covariates, subject to
the constraint that $\alpha(\boldsymbol{c}) +
u\myp\beta(\boldsymbol{c})>0$. It is easy to verify that the
formulas \eqref{Eq:Thres_stab_nonstat_form_sigm_PROPOSED} secure the
threshold stability condition \eqref{Eq:Thres_stab_nonstat_form}:
\begin{align*}
\sigma_u(\boldsymbol{c}) +x\,\xi(\boldsymbol{c} ) &=
\bigl(\alpha(\boldsymbol{c}) + u\myp
\beta(\boldsymbol{c})\bigr)\,\mathrm{e}^{\gamma(\boldsymbol{c})}+
x\mypp \beta(\boldsymbol{c})\,\mathrm{e}^{\gamma(\boldsymbol{c})}\\
&= \bigl( \alpha(\boldsymbol{c})+(u+x)\mypp
\beta(\boldsymbol{c})\bigr)\,\mathrm{e}^{\gamma(\boldsymbol{c})}=\sigma_{u+x}(\boldsymbol{c}).
\end{align*}

By analogy with the regression setting
\eqref{Eq:param_covariat_functional_form}, it is natural to
introduce the following
\begin{assumption}\label{as2}
The coefficients $\alpha(\boldsymbol{c})$,
$\beta(\boldsymbol{c})$ and $\gamma(\boldsymbol{c})$ of the model
\eqref{Eq:Thres_stab_nonstat_form_sigm_PROPOSED}, as well as $\logit
\rho_{u}(\boldsymbol{c})$ as before, are linear functions of the
covariates,
\begin{equation}\label{Eq:param_covariat_functional_form*}
\alpha(\boldsymbol{c}) =
\boldsymbol{a}^{\top}\tilde{\boldsymbol{c}},\ \ \quad
\beta(\boldsymbol{c}) =
\boldsymbol{b}^{\top}\tilde{\boldsymbol{c}},\ \ \quad
\gamma(\boldsymbol{c}) =
\boldsymbol{g}^{\top}\tilde{\boldsymbol{c}},\ \ \quad \logit
\rho_u(\boldsymbol{c}) =
\boldsymbol{r}_u^{\top}\myp\tilde{\boldsymbol{c}}.
\end{equation}
\end{assumption}
Note that the regression vectors $\boldsymbol{a}$, $\boldsymbol{b}$
and $\boldsymbol{g}$ do not depend on the threshold~$u$, in accord
with the construction of the threshold-stable Model~II. So we need
to estimate these coefficients only once, for a suitable level $u$,
then they can be used for any level $u+x$. As for the threshold
exceedance rate $\rho_u$, once it has been estimated at the level
$u$ it can be directly calculated for another level $u+x$ using the
formula \eqref{eq:rhO_u+x}.

\subsection{Bayesian model estimation with MCMC}
\label{subsec:MCMC}

Recall that MCMC simulation techniques are based on running a
suitable Markov chain whose equilibrium distribution is the desired
(posterior) distribution of the model parameters (Gilks et al.,
1996). In addition to computational convenience and efficiency, this
methodology provides more information as compared to the
conventional maximum likelihood inference; for instance, it makes it
possible to find the predictive distribution for future extreme
events (Beirlant et al., 2004).

Estimation of the parameters in our models was carried out using the
\emph{Metropolis--Hastings MCMC algorithm} (Gilks et al., 1996). The
general framework of this algorithm is as follows. Let
$\boldsymbol{X}$ be a random sample from the target distribution
with density $f(\boldsymbol{x}\myp|\myp\boldsymbol{\theta})$, and
let $q(\boldsymbol{\theta})$ be the prior density of the unknown
vector parameter $\boldsymbol{\theta}$. The likelihood function is
denoted by
$L(\boldsymbol{\theta}\myp|\myp\boldsymbol{X})=f(\boldsymbol{X}\myp|\myp\boldsymbol{\theta})$.
According to the Bayes theorem, the posterior density
$q(\boldsymbol{\theta}\mypp|\myp\boldsymbol{X})$ is proportional to
$L(\boldsymbol{\theta}\myp|\myp\boldsymbol{X})\,
q(\boldsymbol{\theta})$. The aim of the Metropolis--Hastings
algorithm is to generate a sample from the posterior distribution
$q(\boldsymbol{\theta}\mypp|\myp\boldsymbol{X})$ obtained as the
equilibrium distribution $\pi(\boldsymbol{\theta})$ of a certain
Markov chain $(\boldsymbol{\theta}_k)$ in the parameter space
(whereby the next state $\boldsymbol{\theta}_{k+1}$ depends only on
the present state $\boldsymbol{\theta}_k$ but not on the past states
$\boldsymbol{\theta}_j$ with $j<k$). In turn,
$\pi(\boldsymbol{\theta})$ can be approximated by a long run of the
Markov chain $(\boldsymbol{\theta}_{k})$, which relies on good
mixing of the chain and fast enough convergence to the equilibrium.
The suitable Markov chain is implemented using a generalized
\emph{acceptance--rejection} sampling method. Namely, at each step,
given the current state $\boldsymbol{\theta}$ of the chain a new
candidate state $\boldsymbol{\theta}'$  is generated according to a
\emph{proposal density}
$g(\boldsymbol{\theta}'\myp|\mypp\boldsymbol{\theta})$, which is
accepted with probability
\begin{equation}\label{eq:a}
A(\boldsymbol{\theta}'\myp|\mypp\boldsymbol{\theta})=\min\left\{1,\frac{q(\boldsymbol{\theta}'|\myp\boldsymbol{X})
\,g(\boldsymbol{\theta}\myp|\myp\boldsymbol{\theta}')}{q(\boldsymbol{\theta}\myp|\myp\boldsymbol{X})\,g(\boldsymbol{\theta}'
|\myp\boldsymbol{\theta})}\right\}.
\end{equation}

Using \eqref{Eq:GPD - Con.Di.F.} and \eqref{eq:rho_c}, the
likelihood function in our GPD-based models is given by (cf.\ Eastoe
\& Tawn, 2009)
\begin{equation*}
\label{Eq:Full-Model-Likelihood}
L(\boldsymbol{\theta}\myp|\myp\boldsymbol{X})=\prod_{t=1}^{n}
\bigl(1-\rho_u(\boldsymbol{C}_t)\bigr)^{1-\delta_u(X_t)} \left(
\frac{\rho_u(\boldsymbol{C}_t)}{\sigma_u(\boldsymbol{C}_t)}\left(
1+\xi (\boldsymbol{C}_t)\mypp\frac{X_t-u\,
}{\sigma_u(\boldsymbol{C}_t)}\right)_{+}^{-1-1/\xi(\boldsymbol{C}_t)}
\right)^{\delta_u(X_t)} ,
\end{equation*}
where $\boldsymbol{\theta}=(\boldsymbol{s}_u,\boldsymbol{\kappa},
\boldsymbol{r}_u)$ (Model~I, see
\eqref{Eq:param_covariat_functional_form}) or
$\boldsymbol{\theta}=(\boldsymbol{a},\boldsymbol{b},\boldsymbol{g},
\boldsymbol{r}_u)$ (Model~II, see
\eqref{Eq:Thres_stab_nonstat_form_sigm_PROPOSED} and
\eqref{Eq:param_covariat_functional_form*}), and $\delta_u$ is the
threshold exceedance indicator: $\delta_u(X_t)=1$ if $X_t>u$ and
$\delta_u(X_t)=0$ otherwise. The role of $\delta_u$ is to dispatch
the correct contribution of each observation $X_t$ depending on
whether $X_t>u$ or not. If there are $m$ covariates under study
(i.e., $\boldsymbol{c}\in\mathbb{R}^m$ and
$\tilde{\boldsymbol{c}}\in\mathbb{R}^{m+1}$) then the dimension of
the parameter vector $\boldsymbol{\theta}$ is $M=3\myp (m+1)$
(Model~I) or $M=4\myp (m+1)$ (Model~II).

For our purposes, the proposals are sampled from a normal
distribution centred at the current state $\boldsymbol{\theta}$ of
the Markov chain, with a fixed (small) standard deviation tuned in
advance. The proposal density is then symmetric,
$g(\boldsymbol{\theta}'
|\myp\boldsymbol{\theta})=g(\boldsymbol{\theta}\myp
|\myp\boldsymbol{\theta}')$, and therefore cancels out from the
acceptance probability \eqref{eq:a}. The priors for different
parameters are taken to be independent and flat (uninformative),
that is, $q(\boldsymbol{\theta})=1$. As a result, the acceptance
probability \eqref{eq:a} is reduced to
\begin{equation}\label{eq:a*}
A(\boldsymbol{\theta}'\myp|\mypp\boldsymbol{\theta})=\min\left\{1,\frac{L(\boldsymbol{\theta}'\myp|\myp\boldsymbol{X})}
{L(\boldsymbol{\theta}\myp|\myp\boldsymbol{X})}\right\}.
\end{equation}

Thus, our MCMC algorithm runs as follows. \noindent
\begin{enumerate}
\item
Initialize the parameter vector by setting
$\boldsymbol{\theta}_0^\top=(\boldsymbol{0},\boldsymbol{0},\boldsymbol{0})$
(Model~I) or
$\boldsymbol{\theta}_0^\top=(\boldsymbol{a}_0,\boldsymbol{0},\boldsymbol{0},\boldsymbol{0})$
(Model~II), where the vector $\boldsymbol{a}_0$ has components
$(\boldsymbol{a}_{0})_i=0$ for $i=1,\dots,m$ and
$(\boldsymbol{a}_{0})_i=1$ for $i=0$. (Note that under this choice
the condition $\alpha(\boldsymbol{c}) + u\myp
\beta(\boldsymbol{c})>0$ of Model~II is satisfied for all
$\boldsymbol{c}\in\mathbb{R}^m$.)

\item
At each step $k\ge 1$:
\begin{itemize}
\item[(i)] Draw a new proposal $\boldsymbol{\theta}'$ from the
normal distribution
$\mathcal{N}_M(\boldsymbol{\theta}_{k-1},\Sigma)$, that is, centred
at the current value $\boldsymbol{\theta}_{k-1}$ and with a fixed
diagonal covariance matrix $\Sigma$ tuned in advance so as to ensure
an optimal acceptance rate of $30$ to $70\%$ (cf.~(iii)).

\item[(ii)] In Model II, check that the proposal $\boldsymbol{\theta}'$ satisfies the constraint
$\alpha(\boldsymbol{c}) + u\myp \beta(\boldsymbol{c})>0$.

\item[(iii)]  Generate an independent random value $U_k$ with uniform distribution
on $[0,1]$. With
$A(\boldsymbol{\theta}'\myp|\mypp\boldsymbol{\theta})$ defined in
\eqref{eq:a*}:
\begin{itemize} \item[$\bullet$] If $U_k\le
A(\boldsymbol{\theta}'\myp|\mypp\boldsymbol{\theta})$ then accept
the proposal and set $\boldsymbol{\theta}_{k}=\boldsymbol{\theta}'$.
\item[$\bullet$] Otherwise, reject the proposal and keep the current value,
$\boldsymbol{\theta}_{k}=\boldsymbol{\theta}_{k-1}$.
\end{itemize}
\end{itemize}
\item
Reset $k\leftarrow k+1$ and go to~2.
\end{enumerate}

Convergence of the sampling algorithm is monitored by the usual
diagnostic tools including visual inspection of the output plots
(Gilks et al., 1996).

\subsection{Variable selection}

An important issue in statistical analysis of air pollution problems
is the choice of an optimal model, that is, deciding on which of the
$m$ covariates should be included in the model to explain most of
the variation in the responses. In the Bayesian context, this
problem is handled by estimating the posterior probability of all
possible models (O'Hara \& Sillanp\"a\"a, 2009). The standard
procedure (Kuo \& Mallick, 1998) is to embed an indicator vector
$\boldsymbol{I}=\left(I_1,\dots,I_m\right)$ into the model, where
$I_j=1$ if the $j$-th covariate is included and $I_j=0$ otherwise
($j=1,\dots,m$). For simplicity, we choose a flat (uninformative)
prior distribution of $\boldsymbol{I}$, so that its components $I_j$
are mutually independent and symmetric (i.e., $\PP(I_j=1)=0.5$),
making all of the $2^m$ possible models equally weighted. Thus, the
prior distribution of the number of covariates included in the model
is binomial with parameters $m$ and $1/2$. The posterior
distribution of the vector $\boldsymbol{I}$ (also called the
\emph{posterior inclusion probabilities}) measures the
data-supported significance of the various covariates, thus advising
the selection of suitable variables.

To incorporate the variable selection in the MCMC simulation, the
Metropolis--Hastings algorithm described in Section
\ref{subsec:MCMC} should be adapted via retaining at each step a
reduced vector of parameters $\mypp\bar{\myn\boldsymbol{\theta}}_k$
according to nonzero values of the current indicator vector
$\boldsymbol{I}=\boldsymbol{I}_k$, which is in turn simply resampled
at each step in an i.i.d.\ fashion.

\subsection{Covariates}
\label{subsec:Covariates}

Choosing the relevant explanatory covariates to draw inference is a
key step in the model building, upon which the model performance
depends rather strongly. The sensitivity of the model to the impact
of different covariates is pinpointed by the complexity of the
physical and chemical mechanisms governing the pollutant
concentrations (Thompson et al., 2001). The traffic data at our
disposal (see the Introduction) are specified using the
\emph{traffic flow} (\emph{TF} = number of vehicles per 15 min) and
\emph{traffic speed} (\emph{TS} = average speed over 15 min,
measured in kph = kilometres per hour), both monitored for the two
categories, light duty vehicles (LDV) including cars, and heavy
goods vehicles (HGV). In turn, meteorological data are encoded using
the following variables: \emph{relative humidity} (\emph{RH});
\emph{solar radiation} (\emph{SR}); \emph{wind speed} (\emph{WS});
\emph{wind direction} (\emph{WD}); and \emph{temperature}
(\emph{T}).

To account for physical and chemical covariates for which we have no
data (such as temporal patterns of potential point sources, e.g.,
factories and other industrial units displaying seasonal behaviour),
we introduce the so-called \emph{composite variables}. The choice of
these variables is also based on seasonality and periodicity
analysis of the measured concentrations. We use Fourier components
accounting for seasonal/periodic oscillations (i.e., yearly, weekly
and daily). The assumption that the model parameters vary smoothly
with certain covariates is consistent with the underlying physical
and chemical mechanisms. Therefore, instead of using the time
variables as factor or dummy variables, they are converted into
circular ones. Further advantage of this approach is that the number
of parameters for estimation is substantially reduced as compared to
using dummy variables.

A pilot study was conducted using the variable selection technique
(see Section~\ref{subsec:MCMC}) to assess the significance of these
variables. The results suggested that the terms up to the second
order for all circular variables and up to the third order for the
daily Fourier components should be sufficient for inclusion as the
model covariates. In addition, interaction terms between the wind
speed and other meteorological variables were also deployed. Lagged
concentrations were included in the model with the aim to account
for residual dependencies due to unobserved covariates. Based on the
examination of the corresponding partial autocorrelation plots, it
was concluded that the fourth-order autoregressive scheme was
adequate.

Let us also make a few comments about the \emph{traffic variables}.
According to Bell et al.\ (2006), for the average UK vehicle fleet
the $\text{NO}_{x}$ emissions are highest at lowest cruise speeds,
significantly decreasing until speeds reach about 60--70 kph, above
which emissions tend to increase slightly at higher speeds. Once
traffic flows reach the capacity of the road, they become unstable
and congestion can be caused. The resulting acceleration,
deceleration and idling of vehicles within the flow generate
elevated emissions. In contrast, the traffic regime that provides
lowest emissions is driving at steady state (cruise) avoiding
alternating periods of acceleration, deceleration and idling.
Following Bell et al.\ (2006) who categorized the traffic regimes in
urban environments into states corresponding to different levels of
emissions congestion, the observed \emph{traffic regime}
(\textit{TR}) was treated in our study as a factor with four levels:
$0\,{=}\myp\text{``quiet''}$ ($\textit{TF}\le 200$, $\textit{TS}\ge
30$); $1\mypp{=}\myp\text{``free''}$ ($200<\textit{TF}\le 300$,
$\textit{TS}\ge 30$); $2\,{=}\myp\text{``busy''}$
($\textit{TF}>300$, $\textit{TS}\ge 30$); and
$3\,{=}\myp\text{``congested''}$ ($\textit{TS}<30$).

Overall, our models include $7$ traffic-related variables; $18$
composite (Fourier) variables; $15$ meteorological variables; and
$12$ lagged concentration values (i.e., up to $4$ lags per each of
the three air pollutants under study). Thus, in total there are
$m=52$ covariate variables in each model.

\subsection{Estimating extreme events by simulation}
\label{subsec:Est_extr_events_by_sim}

In applications of extreme value theory, predictive inference for
unobserved events requires extrapolation. In the air pollution
context, to predict future extremes of the pollutant concentration
process $(X_t)$ (regardless of the covariate process
$(\boldsymbol{C}_t)$) it is common to calculate the \emph{marginal
return levels} $\ell_p$ (for small $p$), defined as the
$100(1-p)\%$-quantile, that is, $\PP(X_t>\ell_p)=p$. Under the
assumption of stationarity of $(X_t)$, the level $\ell_p$ does not
depend on time $t$ and is exceeded on average about once per $n=1/p$
observations. In environmental applications, the mesh size of the
observation grid may vary (e.g., in our case measurements are taken
every minute and the maxima are registered every 15 minutes), so to
give a meaningful interpretation for marginal returns levels not
depending on the grid, it is conventional to report them with regard
to a certain physical amount of time, usually measured in years. In
our case, for example, a one-year return level would correspond to
$1/p=n=4\,\text{(quarters)}\times 24\,\text{(hours)}\times
365\,\text{(days)}=35\myp{}136$, that is, $p=1/n\approx 2.85\times
10^{-5}$.

Non-stationarity can be handled in a similar manner by using the
\emph{conditional distribution} of $X_t$ given the covariate values
at time $t$; namely, the \emph{conditional return level} $\ell_{p}$
is defined by the relation
\begin{equation}\label{Eq:Cond_ret_level}
\PP(X_t>\ell_{p}\,|\,\boldsymbol{C}_t=\boldsymbol{c})=p.
\end{equation}
Note that now $\ell_p=\ell_p(\boldsymbol{c})$ is in fact
time-dependent through the dependence on the current (variable)
value of the covariate vector $\boldsymbol{C}_t=\boldsymbol{c}$.
When the aim is to quantify the effect of future emission (e.g.,
under a new traffic policy) or a possible climate change, formula
\eqref{Eq:Cond_ret_level} provides a simple measure of how a
particular scenario might affect extreme pollution concentration
levels.

In the Bayesian context, estimation of the marginal and conditional
return levels, as well as evaluation of the posterior probabilities
of extreme events, can be carried out using MCMC-generated posterior
samples for the model parameters. For example, a simple way to
estimate the posterior distribution of the marginal return level
$\ell_p$ is to simulate $N$ observations $X_1,\dots,X_N$ ($N\gg
1/p$) using different (independent) samples of parameters from the
corresponding posterior distribution, and then determine an estimate
$\hat{\ell}_p$ from the empirical tail-distribution function by
setting
$$
\hat{\ell}_p=\inf\left\{x\colon \frac{1}{N}\sum_{i=1}^N {\bf
1}_{\{X_i> x\}}>p\right\}=X_{(Np)},
$$
that is, the $(Np)$-th largest observation (order statistic).

To simulate a sequence of observations $(X_t)$ of length $N$, the
simulation method should be nonparametric with regard to the unknown
distribution of the covariates, for which no assumptions are being
made. Following Eastoe (2009), one approach is to resample with
replacement from the observed covariate values and to use the
corresponding empirical distribution. However, nonstationarity of
the covariates on the temporal scale must be preserved, including
seasonal, weekly and daily trends. Hence, the distribution density
$f_t(\boldsymbol{c})$ of the covariate vector $\boldsymbol{C}_t$
(see~\eqref{Eq:X_ut-unconditional}) needs to be estimated from the
covariate values observed under \emph{similar conditions} (e.g., in
the same part of year, day of week, and hour of day), an obvious
drawback of this being that the usable data would be considerably
reduced. The simple alternative approach adopted in the present
study is to utilize the observed values of the traffic and
meteorological covariates rather than simulate them, which
automatically preserves the existing temporal dependence.

A random sequence of exceedances $(X_t^u)$ can be simulated as
follows, using a fourth-order autoregressive scheme within the
covariates (see Section~\ref{subsec:Covariates}).
\begin{enumerate}
\item
Initialize by setting $X^u_t=X_1$ for $t=1,\dots,4$.
\item
For $t\geq 5$, generate an independent random value $U_t$ with
uniform distribution on $[0,1]$. Then, given the observed covariate
vector $\boldsymbol{C}_t=\boldsymbol{c}$ and the preceding values
$X^u_{t-4},\dots,\allowbreak X^u_{t-1}$:
\begin{itemize}
\item [(i)] If
$\rho_{u}(\boldsymbol{c})>U_t$ then simulate $X^u_{t}$ from the GPD
with parameters $\sigma_u(\boldsymbol{c})$ and $\xi(\boldsymbol{c})$
(see~\eqref{Eq:GPD - Con.Di.F.}).
\item [(ii)]
Otherwise (i.e., if $\rho_{u}(\boldsymbol{c})\le U_t$), $X^u_t$ is
plainly resampled from the empirical distribution of observed
concentration values restricted to the region $\{X>u\}$.
\end{itemize}
\item
Reset $t\leftarrow t+1$ and go to~2.
\end{enumerate}

The empirical distribution based on a large number of simulated
samples $\{X^u_{t}\}$ can then be used to estimate the probabilities
of future extreme events.

\section{Results}
\label{sec:Result}

\subsection{Threshold selection}
The results reported here were obtained by applying Models I and~II
described in Section~\ref{sec:Methodology} to the concentration data
shown in Figure~\ref{fig:Time_Series_concentration}. Measurements
were excluded from the analysis if either the concentration values
or one of the covariates were missing; we assume that any missing
observations occur at random due to an independent cause (e.g., a
machine failure) and thus are noninformative. The simplest approach
to choosing an appropriate threshold (Eastoe \& Tawn, 2009) is to
use standard diagnostic tools designed in the stationary framework,
such as mean residual-life plots (see Beirlant et al., 2004). In our
study, this suggested that empirical $90\%$-quantiles could be used
as suitable thresholds, specifically giving the values (in ppb)
$86.94$ for $\text{NO}$, $49.63$ for $\text{NO}_{2}$, and $42.01$
for $\text{O}_{3}$ (see Figure~\ref{fig:Time_Series_concentration}).

In the presence of nonstationarity, the use of such diagnostic tools
is not fully justified and may be questionable. As an alternative,
more sophisticated threshold choice procedures are available in the
nonstationary context; for example, Northrop and Jonathan (2011)
proposed a method for setting covariate-dependent thresholds using
quantile regression, while Northrop et al.\ (2016) focused on
graphical methods for choosing time-dependent thresholds. Such
methods are not used here, since our primary aim is to model
exceedances of fixed thresholds, which in practice may be directly
associated with the air quality standards or critical medical
levels.

Following the ideas briefly outlined in Section~\ref{subsec:MCMC},
MCMC simulation algorithm was employed to obtain inference about the
parameter estimation by using sufficiently long burn-in periods and
different initial values to guarantee convergence of the sampling
algorithm. In order to ensure independence within the sample,
autocorrelation analysis was applied and approximately every
$100$-th generated value was selected and kept for future inference.
Note that choosing a single ``optimal'' model is not satisfactory,
since this would ignore the model uncertainty. In practice, the
problem is circumvented by the \emph{model averaging}, where the
estimate of each candidate model is weighted via its posterior
probability. This approach is especially useful if several candidate
models show high posterior probabilities (O'Hara \& Sillanp\"a\"a,
2009). Moreover, Madigan and Raftery (1994) showed that the model
averaging is an optimal strategy in the sense that it outperforms
any single model in terms of a general utility function derived from
information theory. Therefore, all the results presented below are
obtained for the averaged model. First, Model~I is discussed in
detail and later, in
Section~\ref{subsec:Threshosl_stab_Bayes_model_selec}, it is
compared with Model~II.

\subsection{Assessing the model fit}\label{sec:fit}

Goodness-of-fit of the models can be assessed by using the
well-known \emph{probability integral transform}, whereby $U=F(X)$
is uniformly distributed on $[0,1]$ if $X$ has a continuous
distribution function $F(x)=\PP(X\le x)$. Let
$w_t=\bar{F}_\GPD(X_t^u;\sigma_u(\boldsymbol{C}_t),\xi(\boldsymbol{C}_t))$
be the fitted GPD value for the observed concentration exceedance
$X_t^u=X_t-u$ above threshold $u$ at time $t$ and the corresponding
values of the covariate vector $\boldsymbol{C}_t$ ($t=1,2,\dots$).
Applying the classic Kolmogorov--Smirnov test, we found that the
test accepted the uniformity of $w_t$'s for each pollutant, with the
corresponding $p$-values $0.32$ ($\text{NO}$), $0.18$
($\text{NO}_{2}$) and $0.52$ ($\text{O}_{3}$).

\begin{figure}
\includegraphics[width=1.00\textwidth]
{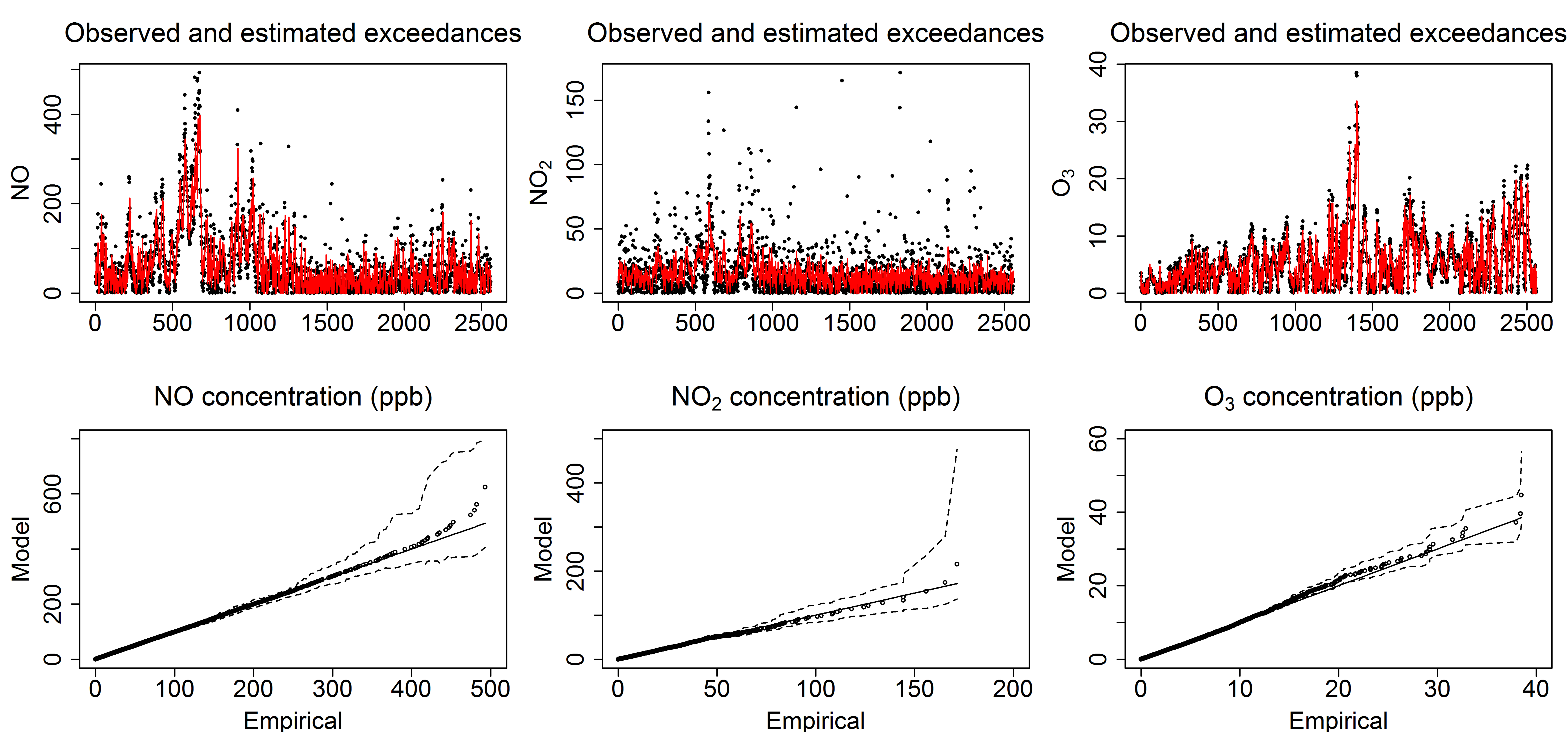} \caption{\textit{Top row:} Observed threshold
exceedances (dots) and their estimated conditional means (solid
lines, red in the online version) for the three pollutants.
\textit{Bottom row:} QQ-plots of the overall model fit to the data,
with straight lines indicating an exact agreement between the model
and observations and dashed lines showing $95$\%-credible intervals.
The data used in these plots were reduced to keep only the
exceedances without missing covariates; the number of usable data
points was $2\myp560$ (out of about $3\myp500$), indexed in the top
row plots in the order of appearance.} \label{fig:QQ_plot_ordered}
\end{figure}
Analysing further the goodness-of-fit of the model, the plots of the
observed threshold exceedances and their estimated posterior means
are shown in the top row of Figure~\ref{fig:QQ_plot_ordered}. (This
figure is based on the data points for which all the covariates were
recorded, resulting in a reduction of usable data from about
$3\myp500$ exceedances down to $2\myp560$.) The expected value of
the GPD \eqref{Eq:GPD -Di.F.} is given by
$\mu^{u}_{\vphantom{t^t}\text{GPD}}=\sigma_u\myp (1-\xi)^{-1}$
($\xi<1$); in the nonstationary context, this is modified
accordingly,
\begin{equation}\label{Eq:GPD_Mean_Exceed_Size_NS}
\mu^{u}_{\vphantom{t^t}\text{GPD}}(\boldsymbol{c})=\frac{\sigma_u(\boldsymbol{c})}{1-\xi(\boldsymbol{c})}.
\end{equation} The median $m^{u}_{\vphantom{t^t}\text{GPD}}=\sigma_u\myp
(2^\xi-1)/\xi$, as well as its nonstationary counterpart
$m^u_{\vphantom{t^t}\text{GPD}}(\boldsymbol{c})$, has an advantage
$\vphantom{\int^t}$ of being well defined for all
$\xi\in\mathbb{R}$, but despite being robust against outliers the
median is more ``conservative'' and may be a less spectacular
benchmark for extreme value models because
$m^{u}_{\vphantom{t^t}\text{GPD}}<\mu^{u}_{\vphantom{t^t}\text{GPD}}$
for $\xi>-1$.

$\vphantom{\int^t}$The plots in Figure~\ref{fig:QQ_plot_ordered}
show that the simulated data reproduce the observed threshold
exceedances reasonably well, which again confirms the overall model
fit. More specifically, for $\text{NO}$ and especially for
$\text{O}_{3}$ the match is quite good, but in the case of
$\text{NO}_{2}$ many threshold exceedances are not well captured by
the mean (nor by the median). The exact reason for that is not quite
clear yet, but it is noteworthy that the $\text{NO}_{2}$ case stands
out as the only one where the estimated shape parameter is
\emph{positive}, $\xi=0.074$ (see Section~\ref{sec:impact}),
suggesting that the posterior GPD of exceedances is rather
heavy-tailed. In this case, the mean is less likely to be a faithful
representative of large exceedances. Furthermore, it may be
misleading to use the standard \emph{coefficient of determination}
$R^2$ as a measure of goodness-of-fit (Draper \& Smith, 1998),
because the coefficient $R^2$ is based on squared deviations from
the mean, which could be extremely large due to heavy tails even if
the model was in fact adequate. As an alternative, the conditional
$95\%$-quantiles were calculated (not presented here), and it was
concluded that the extreme concentrations were captured well by
these higher quantiles. The overall conclusion is that the model has
a good potential for an efficient determination of threshold
exceedances from traffic and meteorological conditions.

Following Eastoe (2009), an alternative way to assess the model's
fit is to plot the observed order statistics of threshold
exceedances against the order statistics obtained by simulation from
the fitted model (cf.\ Section~\ref{subsec:Est_extr_events_by_sim}).
Each of the simulated data sets has the same length as the observed
data. The medians, along with $2.5\%$- and $97.5\%$-quantiles, were
calculated for the order statistics from a simulated data set based
on $1000$ replicas. The results presented in
Figure~\ref{fig:QQ_plot_ordered} (bottom row) also confirm the
overall good model fit.

\subsection{Assessing the impact of the
covariates}\label{sec:impact} At the next step, the interest is in
understanding the qualitative influence of the traffic and
meteorological conditions on the magnitude and occurrence
probability of extreme concentration values. This can be assessed by
analysing the estimated parameters of the GPD as well as the
conditional threshold exceedance probabilities. An optimal threshold
probability that should be used to classify concentration values as
extremes (i.e., lying above the threshold) can be determined by
minimizing the total number of misclassified cases, that is, the
total number of non-extremes classified as extremes and vice versa.
The corresponding misclassification rates were found to be $7.13\%$
($\text{NO}$), $8.26\%$ ($\text{NO}_{2}$) and $7.45\%$
($\text{O}_{3}$). According to these results, the model is adequate
to estimate the time occurrence of the threshold exceedances;
therefore, it can be applied to describe the episodes of elevated
air pollution concentrations.

In view of the equations \eqref{Eq:param_covariat_functional_form},
the average size of exceedances $X^u_t$
(see~\eqref{Eq:GPD_Mean_Exceed_Size_NS}) is a monotone
transformation of linear combinations of the covariates. Hence, the
coefficients of the regression model
\eqref{Eq:param_covariat_functional_form} characterize the size and
direction of the impact of each variable on the threshold
exceedances $X^u_t$. A similar interpretation is valid for the
parameters describing the link between the covariates and the
threshold exceedance rate $\rho_{u}(\boldsymbol{c})$.

Table~\ref{tab:Est_parameters} presents the results of estimation of
the regression coefficients of various covariates for the averaged
model, including the posterior medians, $95\%$-credible intervals
and the posterior inclusion probabilities (note that the latter were
consistently bigger than $0.05$ for almost all the covariates except
for some higher-order Fourier terms of time variables).

\begin{table}
\begin{center}
\caption{Estimated posterior medians and $95\%$-credible intervals
of the regression coefficients (magnified by a factor of $10^{4}$)
for the scale parameter $\sigma_{u}$ and the threshold exceedance
rate $\rho_{u}$ in the
model~\eqref{Eq:param_covariat_functional_form}. The threshold level
$u$ is chosen according to an empirical $90\%$-quantile. The
posterior inclusion probabilities are presented in parentheses.}

\bigskip \label{tab:Est_parameters} {\footnotesize \tabcolsep=0.28pc
\begin{tabular}{l|ccc|ccc} \hline
&&&&&&\\[-.9pc]
\smash{\raisebox{-.51pc}{Covariates}} &\multicolumn{3}{c|}{Scale
parameter $\sigma_{u}$}
&\multicolumn{3}{c}{Threshold exceedance rate $\rho_{u}$}\\[-.1pc]
&$\text{NO}$ &$\text{NO}_{2}$ & $\text{O}_{3}$ &
$\text{NO}$& $\text{NO}_{2}$& $\text{O}_{3}$ \\
&&&&&&\\[-1.2pc]
\noalign{\smallskip}\hline\noalign{\smallskip} \emph{TF}\,(LDV) &
$6.3$ $(0.25)$ & $4.7$ $(0.47)$ &$-2.8$ $(0.25)$ &$37.4$ $(0.91)$&
$38.2$ $(0.86)$&$-13.9$ $(0.88)$\\[-.1pc]
 & $[5.6, 7.7]$ & $[3.5, 5.8]$ &$[-3.2, -2.1]$ &$[32.2, 43.1]$& $[31.5, 45.9]$&$[-18,
 -9.5]$\\[.1pc]
\emph{TF}\,(HGV) & $10.5$ $(0.37)$ & $3$ $(0.33)$ &$-1.9$ $(0.39)$ &$58.5$ $(0.84)$& $33.7$ $(0.79)$&$-20$ $(0.90)$ \\[-.1pc]
 & $[8.4, 12.0]$ & $[2.0, 5.7]$ &$[-2.1, -1.2]$ &$[34.2, 69.7]$& $[22.5, 42.0]$&$[-25.0, -15.6]$\\[.1pc]
\emph{TS}\,(LDV) & $-5.0$ $(0.25)$ & $-6.4$ $(0.78)$ &$1.7$ $(0.13)$ &$-6.7$ $(0.67)$& $-1.7$ $(0.89)$& $6.1$ $(0.55)$ \\[-.1pc]
 & $[-5.2, -4.6]$ & $[-7.6, -5.2]$ &$[1.3, 2.1]$ &$[-9.5, -1.3]$& $[-2.4, -0.7]$&$[5.23, 8.9]$\\[.1pc]
\emph{TS}\,(HGV) & $-1.7$ $(0.27)$ & $-2.5$ $(0.31)$ & $0.79$ $(0.37)$ & $-7.2$ $(0.79)$ & $-1.0$ $(0.12)$& $4.1$ $(0.69)$ \\[-.1pc]
&$[-1.9, -1.5]$& $[-3.8, -1.2]$&$[0.4, 1.2]$ &$[-10.5, -2.8]$& $[-1.6, -0.5]$&$[3.3, 4.9]$\\[.1pc]
\emph{WS} &$-3.1$ $(0.83)$      & $4.4$ $(0.63)$  &$1.9$ $(0.80)$
 &$-4.2$ $(0.90)$ & $-2.0$ $(0.10)$  &$9.3$ $(0.90)$ \\[-.1pc]
 & $[-3.5, -2.6]$ & $[3.9, 6.1]$ &$[1.6, 2.3]$ &$[-5.0, -3.7]$& $[-2.9, -1.4]$&$[4.8, 15.6]$\\[.1pc]
\emph{RH} & $2.3$ $(0.70)$ & $1.6$ $(0.30)$ &$-1.2$ $(0.30)$
&$0.15$ $(0.90)$   & $3.0$ $(0.86)$ &$-1.5$ $(0.44)$ \\[-.1pc]
 & $[2.0, 2.6]$ & $[1.0, 2.8]$ &$[-1.5, -1.0]$ &$[0.1, 0.2]$& $[2.1, 4.2 ]$ &$[-1.9 , -1.0]$\\[.1pc]
\emph{T} & $-1.6 $ $(0.27)$& $-1.7$ $(0.53)$  &$2.6$ $(0.28)$  &$-3.3$ $(0.78)$&$-2.58$ $(0.78)$ &$1.3$ $(0.15)$\\[-.1pc]
 & $[-1.8, -1.6]$ & $[-2.3, -0.9]$ &$[1.9, 3.2]$ &$[-3.6, -2.8]$& $[-3.4, -1.9]$&$[0.9, 1.96]$\\[.1pc]
\emph{SR} & $-3.5$ $(0.62)$ & $-1.3$ $(0.38)$ & $4.0$ $(0.51)$  &$-0.9$ $(0.93)$& $-2.0$ $(0.93)$&$3.9$ $(0.81)$  \\[-.1pc]
 & $[-4.2, -3.5]$ & $[-2.6, -0.7]$ &$[1.8, 6.7]$ &$[-1.4, -0.5]$& $[-2.6, -1.5]$&$[2.9, 4.5]$\\
\noalign{\smallskip}\hline
\end{tabular}}
\end{center}
\end{table}

Note that with a higher traffic flow for both LDV and HGV
categories, the size and probability of threshold exceedances
increase for $\text{NO}_{x}$ and decrease for $\text{O}_{3}$. The
traffic speed has an opposite effect on the size and exceedance
probability for each pollutant, which can be explained by the
negative correlation between the traffic speed and traffic flow. In
practice, the speed is not entirely negatively correlated with the
flow; a positive correlation may arise when the traffic conditions
change from ``busy'' to ``congested'' (Bell et al., 2006). According
to the estimated coefficients of the traffic regimes (factors), the
size and probability of threshold exceedances increase as the
traffic changes from ``quiet'' towards ``congested'', in agreement
with Bell et al.\ (2006).

From the estimated weekly Fourier components of the  time variables
(not shown in the table), there are decreased levels of
$\text{NO}_{x}$ and an increased level of ozone at weekends, as
could be expected due to decreased traffic volume and emission of
possible point sources (e.g., factories). On the other hand,
$\text{NO}_{x}$ levels decrease with temperature and solar
radiation, which is a manifestation of the photochemical nature of
the chemical reaction between these pollutants. The estimated
coefficients of daily Fourier component variables, and their
second-order terms, suggest that there is a significant difference
between daylight and night-time levels, as well as between peak and
off-peak hours, with higher vs.\ lower average concentrations,
respectively. Increased $\text{NO}_{x}$ (decreased $\text{O}_{3}$)
concentrations can be observed during the winter months, contrasted
with decreased $\text{NO}_{x}$ (increased $\text{O}_{3}$)
concentrations during the summer months. Furthermore, stronger winds
correspond to decreased levels of $\text{NO}$ and increased levels
of $\text{NO}_{2}$ and $\text{O}_{3}$, which is likely to be due to
the transport mechanisms and mixing of the particles at this
particular site.

Model~II preserving the threshold stability was fitted to the data
using the similar MCMC procedure as for Model~I (see
Section~\ref{subsec:MCMC}). It is assumed that the parameters
$\alpha(\boldsymbol{c})$ and $\gamma(\boldsymbol{c})$ are linear
functions of the same covariates as in Model~I, while
$\beta(\boldsymbol{c})\equiv\beta$ is constant (see equations
\eqref{Eq:Thres_stab_nonstat_form_sigm_PROPOSED}). The sign of the
shape parameter $\xi$ for different pollutants (determined by the
sign of $\beta$) is the same as in Model~I, that is, the
corresponding GPD is bounded for $\text{NO}$ and $\text{O}_{3}$, and
is heavy-tailed for $\text{NO}_{2}$. The results obtained for the
estimated shape parameter $\xi$ are shown as part of
Table~\ref{tab:shapes}, and are very similar to those in Model~I.
Note however that the estimated medians for $\xi$ are slightly lower
for Model~II as compared to Model~I, so that the former is more
conservative in predicting high exceedances. This trend is also
confirmed by computing the estimated return levels (see
Table~\ref{tab:Return_levels}).


\begin{table}
\begin{center}
\parbox{0.75\textwidth}{\caption{Estimated posterior medians
for the shape parameter $\xi$ and the corresponding $95\%$-credible
intervals (CI) in Models~I and II.} \label{tab:shapes}

\bigskip
\begin{tabular}{c|cc|cc} \hline
&&&&\\[-.9pc]
\smash{\raisebox{-.5pc}{Pollutant}} &
\multicolumn{2}{c|}{\hspace{-2.5pc}\emph{Model I}}
&\multicolumn{2}{c}{\hspace{-2.5pc}\emph{Model II}}\\[.1pc]
\cline{2-5}
&&&&\\[-.9pc]
&$\xi$ & $95\%$-CI&$\xi$ & $95\%$-CI\\[.1pc]
\hline
&&&&\\[-.9pc]
\text{NO}&$-0.101$ & $[-0.114,-0.081]$&$-0.148$&$[-0.156,-0.127]$\\
\ $\text{NO}_{2}$&$\hphantom{-}0.074$ &$[0.038,0.103]$&$\hphantom{-}0.061$&$[0.043,0.091]$\\
$\text{O}_{3}$&$-0.279$ &
$[-0.296,-0.259]$&$-0.292$&$[-0.301,-0.274]$\\[.2pc]
\hline
\end{tabular}

}
\end{center}
\end{table}

One can also plot the estimated medians on top of the observed
exceedances, as was done for Model~I in
Figure~\ref{fig:QQ_plot_ordered}. The results (not shown here) are
very similar, but again suggest that the estimates under Model~II
tend to be slightly lower. The results of estimation of the shape
parameter $\xi$ in Models~I and~II are presented in
Table~\ref{tab:shapes}. In particular, it appears that the posterior
distribution of $\xi$ is likely to be bounded above for $\text{NO}$
and $\text{O}_{3}$ (Weibull type GPD), while being heavy-tailed for
$\text{NO}_{2}$ (Fr\'echet type GPD).

\subsection{Cross-validation}\label{subsec:cross_valis}

The possibility of overfitting as well as the predictive strength of
the model were investigated by using cross-validation. Each month
was divided into two parts\,---\,the first $75\%$ of the data were
used for the model calibration and the remaining $25\%$ for the
model validation. Inference about the predictive strength of the
model was drawn by using the same diagnostics presented in the
previous sections. The results are very similar to those presented
in Figure~\ref{fig:QQ_plot_ordered}, and therefore are not shown
here.

In the prediction of the conditional probability of threshold
exceedance, the corresponding misclassification rates were $10.23\%$
($\text{NO}$), $11.01\%$ ($\text{NO}_{2}$), and $10.79\%$
($\text{O}_{3}$). These results show that the model predicts extreme
concentrations and their occurrence probability quite successfully,
and the risk of overfitting is low.

\subsection{Bayesian model comparison}\label{subsec:Threshosl_stab_Bayes_model_selec}

Although a bare-eye inspection of the graphical results does not
reveal much difference between the models, they can be compared
quantitatively using the so-called \emph{Bayes factor} (Kass \&
Raftery, 1995), defined as the likelihood ratio for Model~II
($\text{M}_2$) against Model~I ($\text{M}_1$),
\begin{equation*}
B_{21}:=\frac{L(\text{M}_2\myp|\myp\boldsymbol{X})}{L(\text{M}_1\myp|\myp\boldsymbol{X})},
\end{equation*}
where $\boldsymbol{X}$ denotes the threshold exceedance data and
$L(\text{M}_i\myp|\myp\boldsymbol{X})=f(\boldsymbol{X}\myp|\myp\text{M}_i)$
is the likelihood (joint density) under the model $\text{M}_i$. The
reference value of the Bayes factor is $B_{21}=1$, corresponding to
no preference for either of the models, while values greater or
smaller than $1$ show evidence in favour of the model $\text{M}_2$
or the alternative model $\text{M}_1$, respectively. It is more
convenient to work with the quantity $\beta_{21}=2\ln B_{21}$, for
which an indicative scale was given by Kass and Raftery (1995); for
example, the values of $\beta_{21}$ in the range 2--5 or 5--10 are
interpreted respectively as \emph{positive} or \emph{strong}
evidence in favour of the model $\text{M}_2$ against the competing
model $\text{M}_1$.

For our data, the calculated Bayes log-factor $\beta_{21}$ is $6.28$
($\text{NO}$), \,$5.57$ ($\text{NO}_{2}$), and $7.49$
($\text{O}_{3}$); thus, according to the above classification the
threshold-stable model $\text{M}_2$ strongly outperforms the
Davison--Smith type model $\text{M}_1$.

This result was also validated and confirmed using an alternative
model selection procedure called the \emph{deviance information
criterion (DIC)} (results not shown here). This criterion is widely
used in the Bayesian setting to handle complex models with abundance
of possible parameters (see Claeskens \& Hjort, 2008).

\subsection{The models' performance under decreased traffic flow}\label{subsec:Dercreased_traffic_folow}

To illustrate the flexibility of the models, possible future
scenarios corresponding to a $25\%$-decrease in the traffic flow
were generated and the drop in the marginal return level was
assessed. Note that the simulation method described in
Section~\ref{subsec:Est_extr_events_by_sim} cannot be used without
modification; indeed, resampling the lagged concentration values
directly from the empirical distribution of the values below the
threshold (i.e., $X_t\le u$) from the same daylight (7am--8pm) or
night-time (8pm--7am) of the corresponding day would lead to a bias
because the reduction in the traffic flow would be ignored. To
overcome this problem, the past concentration values were resampled
from the subset of values $(X_t)$ satisfying $X_t\le u$ and
belonging to the time period of the corresponding daylight or
night-time when the traffic flow is between $70\%$ and $80\%$ of the
total traffic flow observed at time~$t$. If this interval was empty
then the observed lagged concentration values corresponding to time
$t$ were kept.

\begin{table}
\begin{center}
\caption{Estimated 5- and 10-year return levels for the observed and
$25\%$-decreased traffic flow conditions in Models I and~II. Point
estimates and posterior $95\%$-credible intervals were obtained by
simulation.}\label{tab:Return_levels}

\medskip
{\small \tabcolsep=0.40pc
\begin{tabular}{ccccc}
\hline\noalign{\smallskip} &\multicolumn{2}{c}{Observed flow}
&\multicolumn{2}{c}{Decreased flow}\\[-.1pc]
&5-year &  10-year   & 5-year  & 10-year\\[.1pc]
\hline\noalign{\smallskip} \multicolumn{5}{c}{\emph{Model I}}\\
\hline\noalign{\smallskip}
$\text{NO}$ & $988.4$ &  $1264.1$    & $882.0$ & $1055.5$\\[-.1pc]
 & $[591.2, 1325.9]$ &$[780.6,1625.4]$& $[536.8, 1151.5]$ & $[669.3, 1543.9]$\\
$\ \,\text{NO}_{2}$ & $362.3$ &  $426.3$   & $313.1$ & $376.9$\\[-.1pc]
 & $[251.1, 530.5]$ &  $[291.8,602.5]$   & $[216.0, 466.6]$ & $[253.4, 529.6]$\\
$\text{O}_{3}$ & $94.5$ & $99.6$   & $96.7$ &$102.4$\\[-.1pc]
 & $[86.9, 107.9]$ &  $[89.6, 110.5]$   & $[89.1, 111.8]$ & $[93.7, 115.0]$\\
\noalign{\smallskip}\hline
\noalign{\smallskip}\multicolumn{5}{c}{\emph{Model II}}\\
\hline\noalign{\smallskip} $\text{NO}$ & $959.2$
&  $1214.6$     & $859.7$ & $1024.7$\\[-.1pc]
 & $[551.9, 1292.5]$ &$[763.9,1583.9]$& $[519.8, 1139.4]$ & $[652.6, 1517.7]$\\
$\ \,\text{NO}_{2}$ & $336.6$ &  $380.9$   & $288.8$ & $331.1$\\[-.1pc]
 & $[237.8, 499.3]$ &  $[277.7, 566.4]$   & $[238.3, 334.3]$ & $[246.5, 467.6]$\\
$\text{O}_{3}$ & $91.3$ & $97.1$  & $97.0$ & $104.1$\\[-.1pc]
 & $[83.8, 102.9]$ &  $[85.8, 107.4]$   & $[91.5, 114.9]$ & $[94.1, 117.1]$\\
\noalign{\smallskip}\hline
\end{tabular}}
\end{center}
\end{table}

Table \ref{tab:Return_levels} illustrates the $5$- and $10$-year
marginal return levels estimated with the original covariates, as
well as with those corresponding to a reduced traffic flow. Note
that there is a noticeable drop in return levels for each scenario
under Model~II as compared to Model~I. To assess if these estimates
are realistic, additional concentration values were analysed for
cross-validation. Altogether, two years worth of concentration data
were available from the monitoring site covering the period from
November 1, 2007 till November 1, 2009; unfortunately, only the
period from January 1, 2008 to January 1, 2009 could be used for
modelling purposes due to missing traffic or meteorological
observations outside this period.

The observed maxima for the $2$-year data were $811.1$
($\text{NO}$), $245.2$ ($\text{NO}_{2}$), and $80.5$
($\text{O}_{3}$), whereas the observed $1$-year maxima were $574.8$,
$220.3$, and $80.5$, respectively. Overall, these results provide
additional evidence that the estimated 5-year return levels can be
considered as realistic.

\section{Discussion and conclusions}\label{sec:TSummary}

Modelling and predicting air pollution episodes is commonly believed
to be a formidable task, especially on a short time scale. The
inherent difficulties are due to the complexity and dynamic nature
of traffic conditions and meteorological characteristics interwoven
with photochemical reactions in the atmosphere. In this paper, the
classical model of Davison and Smith (1990) was adapted so as to
incorporate nonstationary traffic and weather covariates into the
POT statistical analysis of pollution extremes (Model~I).
Furthermore, a modified version of the Davison--Smith model aiming
to ensure the threshold stability was derived (Model II) and its
performance was compared to that of Model~I. The estimation and a
model selection procedure were carried out using a suitable MCMC
algorithm. Both models demonstrated a good fit to the data; however,
using a Bayesian hypothesis testing it was concluded that Model~II
significantly outperformed Model~I.

The models discussed in this paper yield encouraging results and
have a promising potential for an accurate and reliable estimation
of extreme concentrations. Owing to their regression-based
structure, they are easy to implement in practice. Most importantly,
the models can be used to draw predictive inference about extreme
values beyond the observed ranges and, consequently, to design,
validate and evaluate future air pollution scenarios, for example,
resulting from changing patterns in the traffic flow and/or
meteorological conditions. Thus, our models can provide the air
quality decision makers with an effective tool to manage air
pollution problems.

As compared to a similar model developed by Eastoe and Tawn (2009)
in the context of surface level ozone ($\text{O}_{3}$)
concentrations, a required improvement achieved in the present work
is the inclusion of the traffic-related covariates, which turn out
to be significant under both models. Despite confining ourselves to
univariate models for each chemical, it should be noted that the
need to account for multiple pollutants is somewhat addressed in our
paper owing to the use of a regression framework in the MCMC
simulation; hence, due to lagged past values of the pollutant in
question (e.g., NO), the impact of other (correlated) pollutants
(such as $\text{NO}_{2}$ and $\text{O}_{3}$) is implicitly taken
into account. Furthermore, these models may serve as a stepping
stone for a multivariate version of the POT modelling in the air
pollution context (cf.\ Roth et al., 2014), which will be developed
in our forthcoming work. It would also be important to add a spatial
(e.g., regional) dimension to the models, in particular due to the
apparent significance of proximity to point sources such as
factories or road junctions. To this end, data available from the
UK's largest database known as the Automatic Urban and Rural Network
(AURN, 2016) will be instrumental (cf.\ Gyarmati-Szab\'o et al.,
2011).

The issue of choosing a ``correct'' threshold in the nonstationary
POT modelling has attracted a lot of attention due to its paramount
importance for the extreme value inference (Northrop \& Jonathan,
2011;  Northrop \& Coleman, 2014). Following the classic approach by
Davison and Smith (1990), further developed in the environmental
context, for example, by Eastoe and Tawn (2009), we have opted to
work in this paper with a constant threshold (determined by a
certain empirical quantile, e.g., $90\%$) and to model
nonstationarity of the data through dependence on time-varying
covariates. The alternative popular approach is to choose a
time-dependent, data-driven threshold (using some graphical
diagnostics) but to keep the parameters of the GPD constant
(Northrop et al., 2016; Northrop et al., 2017). The latter idea is
appealing, because the exceedance rate of a fixed threshold may
deteriorate due to nonstationarity; thus, it is reasonable to
monitor the estimation performance and adjust the threshold if and
when necessary. Flexibility with the threshold is also attractive in
view of the possible future changes (e.g., in the environmental
standards, driving patterns or climate). From this point of view,
the threshold-stable Model II proposed in this work is conceptually
rather promising. The fact that we do not seem to use its full
potential is somewhat deceiving: as argued in the paper (see
Section~\ref{subsec:Threshold Stability}), the model fitting should
be done only once for a chosen threshold\,---\,should it change, the
GPD parameters are easily recalculated, whereas the exceedance rate
can be computed using formula \eqref{eq:rhO_u+x}. Thus, in a sense,
Model II may serve to bridge the gap between the two alternative
approaches to the threshold selection.

Finally, let us point out that the successful extreme value
modelling also crucially relies on the quality of the data
collected. Air pollution concentrations are known to be
significantly location-dependent as well as highly variable
temporally, with noticeable deviations from stationarity. Pervasive
mobile environmental sensors, developed for example in the MESSAGE
project (MESSAGE, 2009), could provide a cost-effective, accurate
monitoring system. Air quality data from such a grid, combined with
the urban big data (e.g., UBDC, 2017), would be instrumental for
both feeding in and validating the extreme value models developed in
this study and, as a consequence, should prove valuable for
evidence-based air quality decision-making.


\section*{Acknowledgements}
Research of J.~Gyarmati-Szab\'o was partially supported by an EPSRC
Doctoral Training Grant,
the Strategic Fund of the Institute for Transport Studies and a
Postgraduate Research Scholarship of the School of Mathematics
(University of Leeds). The authors gratefully acknowledge the data
collection and processing provided through the projects ``Mobile
Environmental Sensing System Across Grid Environments'' (MESSAGE),
Department for Transport (DfT) Grant SRT\,7/5/7, and ``Pervasive
Mobile Environmental Sensor Grids'', EPSRC Grant EP/E002013/1
(Institute for Transport Studies, University of Leeds). The authors
are grateful to the anonymous referees for the helpful and
constructive comments.

\bibliographystyle{chicago}


\end{document}